\documentclass[12pt]{iopart}
\usepackage[ansinew]{inputenc}
\usepackage[dvips]{graphics}
\usepackage{epsfig}

\begin{document}

\title{The gravitational analogue to the hydrogen atom\\{\small A summer study at the borders of quantum mechanics and general relativity}}

\author{Martin Kober${}^1$, Benjamin Koch${}^{1,2}$, Marcus Bleicher${}^1$}

\address{${}^1$~Institut f\"ur Theoretische Physik, Johann Wolfgang Goethe-Universit\"at, 
Max-von-Laue-Str.~1, 60438 Frankfurt am Main, Germany \\
${}^2$~Frankfurt Institute for Advanced Studies (FIAS), Max-von-Laue-Str.~1, 
60438~Frankfurt am Main, Germany\\}

\begin{abstract}
This article reports on a student summer project performed in 2006 at the University of Frankfurt. 
It is addressed to undergraduate students familiar with the basic principles of relativistic quantum 
mechanics and general relativity. The aim of the project was to study the Dirac equation in curved space time. 
To obtain the general relativistic Dirac equation we use the formulation of gravity as a gauge theory in the first part. 
After these general considerations we restrict the further discussion to the special case of the Schwarzschild metric. 
This setting corresponds to the hydrogen atom, with the electromagnetic field replaced 
by gravity. Although there is a singularity at the event horizon it turns out that a regular solution of the 
time independent Dirac equation exists. Finally the Dirac equation is solved numerically using
suitable boundary conditions. 
\end{abstract}

\maketitle

\tableofcontents

\section{Introduction}

The most important success in the early days of quantum mechanics was
the appropriate description of the hydrogen atom. It is not
exaggerated to say that quantum mechanics was discovered in its final
form in the 1920ies in the attempt to describe the
hydrogen atom which could not be understood in terms of a classical
theory. 

To describe the behaviour of an electron with the help of quantum
mechanics means to solve the time independent Schr\"{o}dinger equation 
(respectively the Dirac equation in the relativistic case) with a
coupled Coulomb potential and thus to find the eigenstates of the
corresponding Hamiltonian. The existence of bound states within an atom has fundamental 
importance since it is the reason for the stability of matter. So it is suggestive to consider a similar
situation, namely the behaviour of an electron within a spherical symmetric gravitational
field. Since gravitation is much weaker than electromagnetism 
(about a factor of $10^{43}$ in case of the mass and the charge of an electron) the
body being the source of the gravitational field must have a huge
mass to lead to an effect of the same order of magnitude as in the
electromagnetic case. The most massive objects known today are black holes.

Although the situation is in a certain sense similar to the case of
the hydrogen atom, to describe gravitation one has to use general
relativity instead of Maxwell theory. Therefore, the problem becomes much more involved because of the
fundamental difference between the framework of electromagnetism and
general relativity. In the case of general relativity one cannot
simply couple a potential to the Hamiltonian but the structure of space time itself is
changed by the presence of a massive body. This leads to a metric tensor $g^{\mu\nu}(x)$ 
which differs from the usual metric of flat Minkowski space time 
$\eta^{\mu\nu}={\rm diag}(+1,-1,-1,-1)$. The structure of space time near a spherical symmetric 
body is described by the Schwarzschild metric. In the case of the Klein-Gordon 
equation the metric appears explicitly in the equation and one has to replace 
the Minkowski metric by the Schwarzschild metric to obtain the correct equation 
adapted to the new situation. The situation is very different in the case of the Dirac equation, 
because the metric tensor does not appear explicitly in the Dirac equation. Besides this the 
Dirac equation does not describe a scalar valued state but a Spinor. I.e.
that we have to incorporate the effect of the new metric on the Spinor
space. So the situation is more difficult than in the case of the
Klein-Gordon equation where the new metric has just to be
inserted. Therefore we will have to derive the general form of the
Dirac equation in curved space time by considering the gauge
formulation of general relativity as mentioned above. This is the subject of the second
section. After this we will explore the special case of the
Schwarzschild metric and look for a suitable representation to get an
equation which is as simple as possible. 

However the changing of the metric must have an effect
similar to the coupling of a potential. Indeed, as will be shown below, the Dirac equation takes the form of the free
Dirac equation with an additional perturbative term in the Hamiltonian when a certain
representation is chosen. It
remains the task to solve the equation. The main reason why it is more
difficult to solve this problem compared to the hydrogen atom is
the singularity at the event horizon in addition to the
singularity at the origin. However, using a series expansion it turns
out that a solution can be found which is regular at the
horizon. Further it can be shown that a certain boundary
condition has to be fulfilled there. Although the 
probability density of the particle does not vanish within the event horizon we
will omit this area. With the help of the above mentioned series
expansion and boundary condition around the horizon we will get an
initial value infinitesimal close to the event horizon from where a
numerical integration procedure can be started. To start this
integration procedure not directly at the event horizon is necessary
because of the possibility to get to a solution which is not regular
there. As in case of ordinary quantum mechanics the solution has 
to approach zero at infinity to represent a regular 
bound state that is normalizable. With this
condition we can find the eigenstates by varying the energy as long as
the wave function does vanish at infinity. The stationary states
have complex energy eigenvalues as we are dealing with a non-hermitian
Hamiltonian. The complexity of the energy eigenvalues is connected to
the instability of the states and with the help of the imaginary part
of the energy it is possible to determine the decay times. However,
an exact prediction of the decay times exceeds the scope of this article. 
A discussion of this topic can be found in ref. \cite{Lasenby:2002mc}. 
The topic of the summer project as presented here is based on refs. \cite{Lasenby:2002mc,Soffel:1977wn,Soffel:1982pm}
and therefore owes many ideas to these papers where the present problems have been 
treated for the first time. A similar situation concerning the Klein-Gordon equation 
has been considered in ref. \cite{Deruelle:1974zy}.

\section{Implementation of general relativity into the Dirac equation}

Since we are going to study a fermion in a gravitational field we have to consider
the coupling of the Dirac equation to gravity.
According to the other fundamental interactions occurring in nature
gravity can also be formulated as a gauge theory. (For an introduction to the idea of gauge theories in general we suggest ref. \cite{[book6]}(Chapter 15).) The difference is that not local invariance under internal symmetric
transformations is demanded but local invariance under Lorentz transformations is
considered.   

\subsection{Notation}

The following symbols will be used: 

\begin{itemize}
\item[] $\psi$ $\hat =$ Dirac Spinor (in coordinate representation).
\item[]$|\Psi\rangle$ $\hat =$ vector in Hilbert space describing an arbitrary quantum state and $\langle\Psi |$ $\hat =$ the corresponding dual vector (refraining from any special representation).
\item[]$\partial_\mu$ $\hat =$ $\frac{\partial}{\partial x^\mu}$.
\item[]$\gamma^\mu$ $\hat =$ Dirac matrices.
\item[]$\Lambda^\nu_\mu$ $\hat=$ Lorentz transformation matrices and $(\Lambda^{-1})^\nu_\mu$ $\hat=$ the corresponding inverse transformation.
\item[]U $\hat =$ arbitrary unitary operator and $U^{\dagger}$ $\hat=$ its hermitean conjugated operator.
\item[]$x^{[ab]}=x^{ab}-x^{ba}$ 
\end{itemize}
Besides we will make the arrangement that
Greek indices refer to arbitrary global coordinates 
and Latin indices refer to the local coordinate system in which the
metric tensor $g_{\mu\nu}$ takes the form of the special metric tensor
of flat Minkowski space time $\eta_{mn}$. If a certain index adopts
a specific value the special local coordinates are denoted with a
bar to distinguish them from arbitrary global coordinates.
As usual natural units will be used which means that $\hbar=c=G=1$.

\subsection{The general shape of the equation} 

Like in case of usual gauge theories one starts with the Lagrangian of the free field equation,
in our case the Dirac equation (first published in ref. \cite{Dirac:1928hu}) 

\begin{equation}
\mathcal{L}=\bar \psi(i\gamma^\mu\partial_\mu-m)\psi.
\label{16}
\end{equation}
In the framework of relativistic quantum mechanics a Lorentz
transformation is represented by an unitary operator which acts
on a state in Hilbert space. 
Infinitesimal transformations are provided by the generators
of the group. The generators of the Lorentz group obey the following Lie Algebra 

\begin{equation}
[\Sigma_{\alpha\beta},\Sigma_{\gamma\delta}]_-=\eta_{\gamma\beta}\Sigma_{\alpha\delta}-\eta_{\gamma\alpha}\Sigma_{\beta\delta}
+\eta_{\delta\beta}\Sigma_{\delta\alpha}-\eta_{\delta\alpha}\Sigma_{\gamma\beta}.
\label{17}
\end{equation}

If we want to analyse the action of Lorentz transformations on quantum mechanical states of fermions described by the Dirac equation we have to consider the Dirac Spinors representation. In this representation the generators are given by  

\begin{equation}
\Sigma_{\mu\nu}=\frac{i}{4}[\gamma_\mu,\gamma_\nu]_-.
\label{18}
\end{equation}
(For an extensive treatment of the theory of group representations in the context of physics including a derivation of Eq. $(\ref{17})$ and the Dirac Spinor representation of the Lorentz group we suggest ref. \cite{[book2]}(sections 2.2-2.4 and section 5.4, resp.).) 

It is a priori clear that the Dirac equation is invariant
under arbitrary global Lorentz transformations, i.e. independent
of the space time point because it was derived from this constraint. 
(An explicit proof of this invariance property can be found in ref. \cite{[book8]}.)
In contrast to this, the well known Dirac Lagrangian is not invariant under
local Lorentz transformations, but these are exactly the ones needed in the case of general relativity. In general relativity flat Minkowski 
space time is replaced by a curved space time (this topic is originally presented in ref. \cite{Einstein:1916vd}). This means that at every point of space time the metric tensor $g^{\mu\nu}$ looks different. If one wants to compare two mathematical objects at different space time points, one first has to transform to the frame in which the metric tensor looks like the metric tensor $\eta^{\mu\nu}$ of flat Minkowski space time. Thus a specific Lorentz transformation at each space time point has to be performed. This is the reason why we have to consider local Lorentz transformations within the framework of general relativity. Thus, it is our next task to formulate the Dirac equation in a way that obeys local symmetry transformations

\begin{equation}
\psi\ \rightarrow\ U(x)\psi\quad,\quad\partial_\mu\ \rightarrow\ \Lambda_\mu^\nu(x)\partial_\nu.
\label{95}
\end{equation}
A Lagrangian of the following shape satisfies invariance under local Lorentz transformations

\begin{equation}
\mathcal{L}=\bar \psi(i\gamma^\mu D_\mu-m)\psi,
\label{25}
\end{equation}
where $D_\mu$ denotes a new derivative operator (covariant derivative) which transforms under local 
Lorentz transformations according to

\begin{equation}
D_\mu\rightarrow\Lambda_\mu^\nu U(x) D_\nu U^{\dagger}(x).
\label{26}
\end{equation}
It is clear that such a covariant derivative obeys local Lorentz invariance,
because the $U$ operators to the left and right of the
derivative operator cancel the operators which act on the right and on
the left state in the Lagrangian.

The covariant derivative $D_\mu$ applied to the metric tensor is
zero. This means that $D_\mu$ is the derivative connecting different 
space time points endowed with a local pseudo-Euclidean coordinate system in which $g_{\mu\nu}=\eta_{\mu\nu}$. Since we are
regarding Spinors on the space time it is also necessary to consider
the action of $D_\mu$ on the Dirac Spinor space.  

Now we have to find out how this covariant derivative looks like. To do this, we assume that $D_\mu$ is of the following shape

\begin{equation}
D_m=e_m^\mu(\partial_\mu+i\omega_\mu).
\label{27}
\end{equation}
Here we have introduced the vierbein $e_m^{\nu}$ describing the transformation between
arbitrary global coordinates and the local coordinates in which the
metric tensor looks like the metric tensor of flat
Minkowski space time. This means that the following relation is valid

\begin{equation}
\eta_{mn}=e_{m}^{\mu}e_{n}^{\nu}g_{\mu\nu},
\label{28}
\end{equation}
which can be seen as the definition of the vierbein. It expresses that if we know the vierbein at each space time point we can obtain the metric field meaning that it contains the same information about the structure of space time. So the vierbein represents another way to describe the gravitational field and can replace the metric tensor.
Further, the vierbein fulfils the relations

\begin{equation}
e_{\mu}^{m}e_{n}^{\mu}=\delta^{m}_{n}\quad , \quad e_m^{\mu}e_{\nu}^{m}=\delta^{\mu}_{\nu}.
\label{29}
\end{equation}
(An extensive introduction to the vierbein-formalism can be found in ref. \cite{[book3]}).

By inserting Eq. $(\ref{27})$ into the above transformation condition Eq. $(\ref{26})$ and rewriting it a little bit one gets

\begin{equation}
e_m^\mu(\partial_\mu+i\omega_\mu)\rightarrow\Lambda_m^n e_n^\mu\left[\partial_\mu+U(x)\partial_\mu U^\dagger (x)+iU(x)\omega_\mu U^\dagger (x)\right].
\label{31}
\end{equation}
So the covariant derivative $e_m^\mu(\partial_\mu+i\omega_\mu)$
obeys condition $(\ref{27})$ if $e^\mu_n$ and $\omega_\mu$ transform according to

\begin{equation}
e^\mu_n \rightarrow \Lambda^n_m e^\mu_n\quad,\quad \omega_\mu\rightarrow U\omega_\mu U^{\dagger}-iU(\partial_\mu U^{\dagger}).
\label{32}
\end{equation}
The matrices $\omega_\mu$ describe the Dirac Spinor connection. 
They can be performed by the above generators of the Lorentz group of Eq. $(\ref{18})$

\begin{equation}
\omega_\mu(x)=\frac{1}{2}\omega_\mu^{mn}(x)\Sigma_{mn}.    
\label{33}
\end{equation}
So we have found the Lagrangian which is invariant under local Lorentz transformations

\begin{equation}
\mathcal{L}=\bar \psi\left[i\gamma^m e_m^\mu\left(\partial_\mu+\frac{i}{2}\omega_\mu^{mn}\Sigma_{mn}\right)-m\right]\psi,
\label{34}
\end{equation}
leading to the Dirac equation in general relativity

\begin{equation}
\left[i\gamma^m e_m^\mu\left(\partial_\mu+\frac{i}{2}\omega_\mu^{mn}\Sigma_{mn}\right)-m\right]\psi=0.
\label{35}
\end{equation}
(For an extensive discussion of gravity as a gauge theory the reader is referred to ref. \cite{[book3]}.)

Combining the well-known anti-commutation relations for the Dirac-matrices $[\gamma^\mu,\gamma^\nu]_+=2\eta^{\mu\nu}$ and $(\ref{28})$ leads to the new anti-commutation relations

\begin{equation}
[e^\mu_m\gamma^m,e^\nu_n\gamma^n]_+=2g^{\mu\nu},
\label{36}
\end{equation}
for the Dirac matrices in curved space time. We will need this relation in the next chapter.\\ 

\subsection{Calculation of the Spinor coefficients} 

Now we have to determine the specific form of the Spinor connection coefficients $\omega_\mu^{mn}$.
It is important to notice that our coefficients $\omega_\mu^{mn}$ contain one index $(\mu)$ referring to the global coordinate system and two indices $(m,n)$ referring to the local coordinate system where $g_{mn}=\eta_{mn}$.
First, we express the above coefficients $\omega_\mu^{mn}$ by the usual
Christoffel symbols $\Gamma^\rho_{\mu\nu}$ defining the parallel transport between vectors 
in the tangential vector space of different space time points. 
They are defined by the condition that the metric shall be left invariant by the covariant derivative 

\begin{equation}
D_\mu (g_{\rho\sigma})=0
\label{38}
\end{equation}

and have the following shape 

\begin{equation}
\Gamma^\rho_{\mu\nu}=\frac{1}{2}g^{\rho\sigma}(\partial_\nu g_{\mu\sigma}+\partial_\mu g_{\nu\sigma}-\partial_\sigma g_{\mu\nu}).
\label{41}
\end{equation}

(The explicit derivation of this relation can be found in any book on general relativity, e.g. \cite{[book4]}.)

We can express the covariant derivative of the vierbein by applying the Christoffel symbol to each index
leading to 

\begin{equation}
D_\mu e^m_\nu=\partial_\mu e^m_\nu+\Gamma^m_{\mu n}e^n_\nu-\Gamma^\rho_{\mu\nu}e^m_{\rho}.
\label{42}
\end{equation}
Inserting Eq. $(\ref{28})$ into Eq. $(\ref{38})$ and considering the case $\nu=\rho$ one obtains

\begin{eqnarray}
D_\mu(g_{\nu\nu})&=D_\mu (e_\nu^m e_\nu^n \eta_{mn})=D_\mu ((e_\nu^m)^2 \eta_{mm})\nonumber\\
&=\eta_{mm} D_\mu (e_\nu^m)^2=2\eta_{mm}e_\nu^m D_\mu e_\nu^m=0.
\label{43}
\end{eqnarray}
which means that the covariant derivative applied to the vierbein gives zero

\begin{equation}
D_\mu e^m_\nu=0.
\label{44}
\end{equation}
Here we have used that $D_\mu\eta_{mn}=0$ and $\eta_{mn}$ is a diagonal matrix. 
Combining Eq. $(\ref{42})$ and $(\ref{44})$ results in  

\begin{equation}
\partial_\mu e^m_\nu+\Gamma^m_{\mu n}e^n_\nu-\Gamma^\rho_{\mu\nu}e^m_{\rho}=0.
\label{45}
\end{equation}
By moving the second term of Eq. $(\ref{45})$ to the r.h.s. side of the equation, 
multiplying with $e^\nu_n$ and raising the flat coordinate indices
with $\eta^{\alpha\beta}$ we arrive at a 
relation between the above Dirac Spinor coefficients and the Christoffel 
symbols in arbitrary coordinates.

\begin{equation}
\omega_\mu^{mn}=\Gamma_{\mu}^{mn}=-\eta^{n p}e^{\nu}_{p}(\partial_\mu e_\nu^m-\Gamma^\rho_{\mu\nu} e_\rho^m).
\label{46}
\end{equation}
Here, we can identify the coefficients of the connection $\omega_\mu^{mn}$ with the Christoffel symbols expressed in terms of the local coordinates, because the connection between the tangential vector spaces of different points on the space time manifold is independent of the special space by which it is represented. If one recalls now that the vierbein $e^\mu_m$ describes the transformation
between global and local coordinates, one can express the coefficients by the vierbein. 
Using Eq. $(\ref{28})$ in Eq. $(\ref{41})$, inserting it to 
Eq. $(\ref{46})$ and transforming the obtained expression a little bit, we arrive at 

\begin{equation}
\omega_\mu^{mn}=2e^{\nu [m}\partial_{[\mu}e_{\nu]}^{n]}+e_{\mu p}e^{\nu m}e^{\sigma n}\partial_{[\sigma}e_{\nu]}^p.
\label{48}
\end{equation}
So we have expressed the connection coefficients in terms of the vierbein. (The derivation of this formula can also be found in \cite{[book1]}.
For advanced aspects concerning the relation between the usual formulation
of general relativity and the vierbein formalism we suggest Refs. \cite{[book3]},\cite{[book1]} and \cite{[book5]}.)

\section{Solution of the Dirac equation in a Schwarzschild background}

\subsection{The choice of a suitable formulation of the Dirac equation in a Schwarzschild background}

So far we have formulated the Dirac equation in a general space time
with an arbitrary metric tensor and an arbitrary connection between
different space time points. To be specific we will from now on focus on the special case of
the Schwarzschild metric, because we are investigating a particle in a
spherically symmetric gravitational field. 

The Schwarzschild metric (first presented in \cite{Schwarzschild:1916uq}) is usually represented in the following way

\begin{equation}
g_{\mu\nu}=\left(\begin{array}{cccc}
(1-\frac{2M}{r}) & 0 & 0 & 0\\
0 & \frac{-1}{(1-\frac{2M}{r})} & 0 & 0\\
0 & 0 & -r^2 & 0\\ 
0 & 0 & 0 & -r^2 {\rm sin}(\theta)\end{array}\right).
\label{49}
\end{equation}
A spherically symmetric object only influences the radial part of the
metric. Therefore, we have to deal with the time and the radial
part of the metric and do not consider the angular coordinate. This leads to

\begin{equation}
\begin{array}{ccc}
e^0_\mu\gamma^\mu &=& e^0_{\bar 0}\gamma^{\bar 0}+e^0_{\bar r}\gamma^{\bar r},\\
e^r_\mu\gamma^\mu &=& e^r_{\bar 0}\gamma^{\bar 0}+e^r_{\bar r}\gamma^{\bar r}.
\end{array}
\label{50}
\end{equation}
The components of the Schwarzschild metric are not fully determined 
due to the fact that the metric can be expressed as a function of a new time coordinate according to a
transformation of the form $t\rightarrow t+x(r)$ where $x$ denotes an
arbitrary function of $r$. It is important to note that this
transformation of the time coordinate 
indeed changes the appearance of the metric, but does not affect the
$g_{00}$-component of the metric. This component is equal in all representations 

\begin{equation}
g_{00}=(1-\frac{2M}{r}).
\label{51}
\end{equation}
Besides this, the vierbein is still under-determined if a special time
coordinate is chosen because of the possibility to apply a radial
dependant Lorentz boost to the system connected to the choice of a global gauge. 

This means that two degrees of freedom remain. However, two constraints can
be formulated. The first constraint can be derived from the fact that the new objects
$e^0_\mu\gamma^\mu$ and $e^r_\mu\gamma^\mu$ have to fulfil the same
Lie-Algebra as $\gamma^0$ and $\gamma^r$ which means that

\begin{equation}
[\gamma^0,\gamma^r]_-=[e^0_m \gamma^m,e^r_n\gamma^n]_-.
\label{54}
\end{equation}
And the second constraint arises from the invariance of the $g_{00}$-component of the metric.
By applying the covariant metric tensor $g_{\mu\nu}$ to Eq. $(\ref{36})$ we get two further equations

\begin{equation}
\begin{array}{ccc}
{[e^{\mu}_m\gamma^m,e^n_\nu\gamma_n]_+} &=& 2\delta^\mu _\nu,\\
{[e_{\mu}^m\gamma_m,e_{\nu}^n\gamma_n]_+} &=& 2g_{\mu\nu}.
\end{array}
\label{55}
\end{equation}
So we can formulate the two following constraints

\begin{equation}
\begin{array}{ccc}
e^0_{\bar 0}e^r_{\bar r}-e^0_{\bar r}e^r_{\bar 0}&=&1,\\
(e_0^{\bar 0})^2-(e_0^{\bar r})^2&=&(1-\frac{2M}{r}).
\end{array}
\label{56}
\end{equation}

It is very important to emphasise that all choices of the vierbein fulfilling the constraints $(\ref{56})$ represent the usual Schwarzschild metric but in different representations.
It turns out that the Dirac equation in a Schwarzschild background
becomes quite simple if we choose the coefficients as 

\begin{equation}
e^0_{\bar 0}=1\quad ,\quad e^0_{\bar r}=0\quad ,\quad e^r_{\bar r}=1\quad ,\quad e^r_{\bar 0}=-\left(\frac{2M}{r}\right)^{\frac{1}{2}}.
\label{57}
\end{equation}
It is left to the reader to show that these choices fulfil the conditions $(\ref{56})$.
The inverse components of the vierbein are determined by Eq. $(\ref{29})$ leading to

\begin{equation}
e^{\bar 0}_0=1\quad ,\quad  e^{\bar r}_0=\left(\frac{2M}{r}\right)^{\frac{1}{2}}\quad ,\quad e^{\bar r}_r=1\quad ,\quad e^{\bar 0}_r=0.
\label{58}
\end{equation}
By the special choice of the vierbein in Eq. $(\ref{57})$, the special form of the
metric is completely determined by Eq. $(\ref{36})$ and Eq. $(\ref{55})$.
To find the specific Dirac equation we have to specify the connection coefficients $\omega_\mu^{mn}$ by inserting the above choice of the vierbein to Eq. $(\ref{48})$. In the expression arising from this most terms vanish and one obtains the following coefficients  
 
\begin{equation}
\omega_0^{\bar 0 \bar r}=-\omega_0^{\bar r \bar 0}=-\frac{3M}{r^2} \quad, \quad  
\omega_r^{\bar 0 \bar r}=-\omega_r^{\bar r \bar 0}=-\frac{3}{2}\left(\frac{2M}{r}\right)^{\frac{1}{2}}\frac{1}{r}.
\label{59}
\end{equation}
The coefficients with equal upper indices vanish because of the antisymmetry.
By inserting Eqs. $(\ref{57})$ and $(\ref{59})$ to the general Dirac Eq. $(\ref{35})$ and using the antisymmetry of the generators and the cooresponding coefficients leading to $\omega_\mu^{mn}\Sigma_{mn}=2 \omega_\mu^{\bar 0 \bar r}\Sigma_{\bar 0 \bar r}$ we get the following expression  

\begin{equation}
(i\gamma^\mu\partial_\mu-m-i\gamma^0\left(\frac{2M}{r}\right)^\frac{1}{2}\partial_r
-\gamma^r\left(-\frac{3}{2}\left(\frac{2M}{r}\right)^\frac{1}{2}\frac{1}{r}\right)\frac{i}{4}[\gamma_0,\gamma_r]_-)\psi=0.
\label{60}
\end{equation}
$\gamma^0$ and $\gamma^r$ obey the same anticommutation relation as $\gamma^0$ and $\gamma^i$ where $i$ runs from 1 to 3. From this we see 
that $[\gamma_0,\gamma_r]_-=-2 \gamma_r \gamma_0$ and further that $\gamma^r \gamma_r=1$. So we find that in this representation the Dirac equation takes the following form

\begin{equation}
(i\gamma^\mu\partial_\mu-m-i\gamma^0\left(\frac{2M}{r}\right)^\frac{1}{2}(\partial_r+\frac{3}{4r}))\psi=0.
\label{61}
\end{equation}

(This result for the Dirac equation in a Schwarzschild background with our special choice of gauging corresponds exactly to the one obtained in \cite{Lasenby:2002mc}). 

\subsection{Reformulation to a system of differential equations}

In the previous section we have derived the Dirac equation on a Schwarzschild background in a suitable representation.
It has the form of the free Dirac equation with an additive radial dependant
perturbation term. Now the task remains to reformulate the
free part of the equation and to express it in spherical coordinates
and then to separate the radial part of the state from the
spherical symmetric one.

First we rewrite Eq. $(\ref{61})$ by expressing the $\gamma$-matrices by
the well-known $\alpha$- and $\beta$-matrices which are defined by the relations
$\gamma_0=\beta$ and $\gamma_k=\beta\alpha_k$ for $k=1,2,3$ and obtain

\begin{equation}
(i\beta\partial_t+i\beta\alpha^k\partial_k-m-i\beta\left(\frac{2M}{r}\right)^\frac{1}{2}(\partial_r+\frac{3}{4r}))\psi=0.
\label{62}
\end{equation}
Rearranging this equation and multiplying it with $\beta$ we get 

\begin{equation}
(i\alpha^k\partial_k)\psi=(-i\partial_t+{\beta}m+i\left(\frac{2M}{r}\right)^\frac{1}{2}(\partial_r+\frac{3}{4r}))\psi=0.
\label{63}
\end{equation}
Since we are finally interested in stable orbits (the energy eigenstates) we separate the time to get the time independent Dirac equation.

Choosing the following ansatz

\begin{equation}
\psi=e^{-iEt}\Psi,
\label{64}
\end{equation}
leads to

\begin{equation}
(i\alpha^k\partial_k)\Psi=(-E+{\beta}m+i\left(\frac{2M}{r}\right)^\frac{1}{2}(\partial_r+\frac{3}{4r}))\Psi.
\label{65}
\end{equation}
We can rewrite this to

\begin{equation}
H\Psi=(H_{Dirac}+H_{INT})\Psi=E\Psi,
\label{88}
\end{equation}
where

\begin{equation}
H_{Dirac}=(-i\alpha^k\partial_k+{\beta}m),
\label{89}
\end{equation}
and 

\begin{equation}
H_{Int}=i\left(\frac{2M}{r}\right)^\frac{1}{2}(\partial_r+\frac{3}{4r}).
\label{90}
\end{equation}
$\Psi$ shall describe an eigenstate to $J^2\ ,J_z$ and $P$ where $J$
describes the norm of the overall angular momentum which means the sum
of the orbital angular momentum $L$ and the Spin $S$, $J_z$ its component
in $z$-direction and $P$ the parity.  

The Parity will be denoted by the quantum number $\omega$ (not to be confused with the Spinor coefficients $\omega_\mu^{mn}$). We set
$\omega=+1$ {\rm for states with parity} $(-)^{J+\frac{1}{2}}$ and
$\omega=-1$ {\rm for states with parity} $(-)^{J-\frac{1}{2}}$.

In the following $y_{LJ}^M$ will describe an eigenstate to $J^2$ and $J_z$ where the parity is given by $(-)^L$
which means that $L$ can adopt the two values $L=J+\frac{1}{2}\omega$ and $\bar L=J-\frac{1}{2}\omega$ and that 
$y_{LJ}^M$ and $y_{\bar LJ}^M$ have opposite parity.

Next we make the following ansatz for $\Psi$  

\begin{equation}
\Psi=\frac{1}{r}\left(\begin{array}{cc}
\Phi y_{LJ}^M\\
i \chi y_{\bar LJ}^M\end{array}\right),
\label{66}
\end{equation}
where $\Phi$ and $\chi$ describe arbitrary radial states.
By following a derivation also used in case of the relativistic hydrogen atom (where the free part of the Dirac equation has to be reformulated in spherical coordinates also and which can be found in Ref. \cite{[book7]} for example) and splitting the two components of the equation into a system of two differential equations we obtain

\begin{eqnarray}
\left[-\partial_r+\frac{w(J+\frac{1}{2})}{r}\right]\Phi &=\left[E-m-i\left(\frac{2M}{r}\right)^\frac{1}{2}(\partial_r-\frac{1}{4r})\right]\chi, \nonumber\\
\left[\partial_r+\frac{w(J+\frac{1}{2})}{r}\right]\chi 
&= \left[E+m-i\left(\frac{2M}{r}\right)^\frac{1}{2}(\partial_r-\frac{1}{4r})\right]\Phi.
\label{75}
\end{eqnarray}

These are the differential equations for a Dirac particle in a Schwarzschild metric we have to solve.   

\subsection{Solution of the time independent equation with suitable boundary conditions}

After deriving the system of differential equations the final task is to calculate the energy eigenstates by solving the set of coupled partial differential equations. Because the set of equations contains complex numbers it is convenient to split the two
radial components $\Phi$ and $\chi$ in a real and imaginary
component resulting in four equations which can be integrated
numerically.

However, before we do this, we have to make sure that there exists a
regular solution at the event horizon and to
determine the necessary boundary conditions. Therefore we will
transpose the above equation system. In both Eqs. $(\ref{75})$ appear
the derivatives of $\Phi$ and $\chi$. By solving one equation for
$\partial_r \Phi$ and the other 
for $\partial_r \chi$ and inserting each equation into the other we get the following system

\begin{eqnarray}
\left(1-\frac{2M}{r}\right)\partial_r\Phi&=\left[\frac{w(J+\frac{1}{2})}{r}+i\left(\frac{2M}{r}\right)^\frac{1}{2}[E+m+i\left(\frac{2M}{r}\right)^\frac{1}{2}\frac{1}{4r}]\right]\Phi\nonumber\\
&\quad+\left[i\left(\frac{2M}{r}\right)^\frac{1}{2}[-\frac{w(J+\frac{1}{2})}{r}-\frac{1}{4r}]-(E-m)\right]\chi,\nonumber\\
\left(1-\frac{2M}{r}\right)\partial_r\chi&=\left[-\frac{w(J+\frac{1}{2})}{r}+i\left(\frac{2M}{r}\right)^\frac{1}{2}[E-m+i\left(\frac{2M}{r}\right)^\frac{1}{2}\frac{1}{4r}]\right]\chi\nonumber\\
&\quad+\left[i\left(\frac{2M}{r}\right)^\frac{1}{2}[-\frac{w(J+\frac{1}{2})}{r}+\frac{1}{4r}]+(E+m)\right]\Phi.
\label{76}
\end{eqnarray}
After this transformation the singularity at the event horizon $(r_s=2M)$ shows up explicitly. Nevertheless it is possible to find a regular solution. To obtain this solution we perform a series expansion around the event horizon

\begin{equation}
\Phi=\sum_{i=1}^{\infty} a_i(r-2M)^i\quad,\quad\chi=\sum_{i=1}^\infty b_i(r-2M)^i.
\label{77}
\end{equation}
By inserting now the series expansion into Eq. $(\ref{76})$ we see that it is a
solution of our differential equation system at the event horizon $r=2M$
under condition that

\begin{eqnarray}
\left[iE-\frac{1}{8M}+\left(\frac{\omega(J+\frac{1}{2})}{2M}+im\right)\right]a_0&\nonumber\\
\quad=\left[E+i\frac{1}{8M}+\left(i\frac{\omega(J+\frac{1}{2})}{2M}-m\right)\right]b_0&.
\label{78}
\end{eqnarray} 
Here we have used 
\begin{equation}
\sum_{i=1}^{\infty} x_i(r-2M)^i\vert_{r=2M}=x_0,
\label{79}
\end{equation}
and 
\begin{eqnarray}
\left(1-\frac{2M}{r}\right)\partial_r\sum_{i=1}^{\infty} x_i(r-2M)^i\vert_{r=2M}&\nonumber\\
\quad=\frac{r-2M}{r}\sum_{i=1}^{\infty} ix_i(r-2M)^{(i-1)}\vert_{r=2M}\nonumber\\ 
\quad=\frac{1}{r}\sum_{i=1}^{\infty} ix_i(r-2M)^i\vert_{r=2M}=0.
\label{80}
\end{eqnarray}
To begin an integration of the differential equation system Eq. $(\ref{76})$ one
needs the start values of $\Phi$ and $\chi$. By using the series
expansion Eq. $(\ref{77})$ we have found one constraint for the start values 
of the integration of $\phi$ and $\chi$. The remaining degree of
freedom is only connected to the the overall normalisation of the wave
function. It is determined by the normalisation condition that the
integral over the squared norm has to be equal to unity because of the
probability interpretation of the wave function. So the wave is
defined apart from a physically irrelevant overall phase. 

The integration cannot be started directly from the horizon because of
the problem with the non-regular solutions. 
However, if we consider our series expansion Eq. $(\ref{77})$ to the first order
infinitesimal adjacent to the event horizon we can find the
coefficients
$a_1$ and $b_1$ by inserting it to the equation system $(\ref{76})$ and solving
it. Because of the complexity of $a_1$ and $b_1$ we have effectively
four equations. The calculation is quite long but mathematically
trivial. We will therefore refrain from presenting it here.

The coefficients $a_1$ and $b_1$ of the first order series expansion Eq. $(\ref{77})$  
together with the start values of $\phi$ and $\chi$ define the wave function infinitesimally close to the
horizon from where the numerical integration can be started. The energy eigenvalues are found by the condition that the
wave function has to approach zero at infinity to assure integrability. 

\subsection{Non-hermiticity of the Hamiltonian}

Unfortunately, we cannot be sure that the eigenvalues of the Hamiltonian in Eq. $(\ref{88})$ $H=H_{Dirac}+H_{Int}$ are
real. In general real eigenvalues are only obtained from hermitean operators. In the case discussed here, we will show that the Hamiltonian is not hermitian. 
The Hamiltonian is the generator of the group describing the way quantum mechanical states in Hilbert space evolve with time. This means that

\begin{equation}
U(t_2,t_1)={\rm exp}\left[-iH(t_2-t_1)\right],
\label{91}
\end{equation}
where $U(t_2,t_1)$ describes the relation between a quantum mechanical state $| \psi (t_1) \rangle$ at time $t_1$ and the same state $| \psi (t_2) \rangle$ at time $t_2$ within the Schr\"{o}dinger picture, i.e.

\begin{equation}
| \psi (t_2) \rangle=U(t_2,t_1) | \psi (t_1) \rangle.
\label{92}
\end{equation}
If the time evolution operator $U(t_1,t_2)$ is unitary this results in a constant overall probability. But Eq. $(\ref{91})$ tells us that this is only the case because of the hermiticity of the Hamiltonian.
In the case of a non-hermitean Hamiltonian unitary breaks down and the norm of the state can change. 
In our case this is due to the fact that the singularity at the centre of the black hole leads to
an instability of the states resulting in a probability current through the event horizon. 
Before drawing our attention to this phenomenon again we will first show this non-hermiticity of the Hamiltonian of Eq. $(\ref{88})$.

Hermiticity implies that the expectation value of an operator is real

\begin{equation}
\langle \Psi | H | \Psi \rangle = \langle \Psi | H | \Psi \rangle^{*}.
\label{81}
\end{equation}
In coordinate representation the inner product of Dirac Spinors is defined as 

\begin{equation}
\langle \Phi | \Psi \rangle = \int\limits_{-\infty}^{\infty}\phi^{\dagger}\gamma^0 \psi\ d^3r = \int\limits_{-\infty}^{\infty}\bar \phi \psi\ d^3r.
\label{82}
\end{equation}
As the free Dirac Hamiltonian $H_{Dirac}$ is hermitian, we have to check the additional perturbation Hamiltonian $H_{Int}$

\begin{equation}
\langle \Psi | H_{Int} | \Psi \rangle = \int\limits_{-\infty}^{\infty}\psi^{\dagger}\gamma^0 (-i\gamma^0\left(\frac{2M}{r}\right)^\frac{1}{2}(\partial_r+\frac{3}{4r}))\psi\ d^3r
\label{83}
\end{equation}

\begin{equation}
\langle \Psi | H_{Int} | \Psi \rangle^{*} = \int\limits_{-\infty}^{\infty}\psi^{T}\gamma^0 (i\gamma^0\left(\frac{2M}{r}\right)^\frac{1}{2}(\partial_r+\frac{3}{4r}))\psi^{*}\ d^3r.
\label{84}
\end{equation}
If we integrate the derivative term by parts and use the condition
that the norm of the wave function vanishes at the boundary we
find the expression 

\begin{eqnarray}
\langle \Psi | H_{Int} | \Psi \rangle^{*} &=& 4\pi \int\limits_{0}^{\infty}\left(\left(\partial_r\psi^{T}\right)
\left(-i\left(\frac{2M}{r}\right)^\frac{1}{2}\right)\psi^{*}+\psi^{T}\left(i\left(\frac{2M}{r}\right)^\frac{1}{2}\frac{3}{4r}\right)\psi^{*}\right)r^2\ dr\nonumber\\
&&+4\pi\ \psi^{T} \left(i\left(2Mr^3\right)^\frac{1}{2}\right)\psi^{*} \mid_{r=0}+4\pi \int\limits_{0}^{\infty}\psi^{T}\left(-i\partial_r \left(2Mr^3\right)^\frac{1}{2}\right)\psi^{*}\ dr.\nonumber\\
\label{85}
\end{eqnarray}
Here we have used that $(\gamma_0){}^2=1$. If we now use that $\phi^{T}\psi=\psi^{T}\phi$ we obtain 

\begin{eqnarray}
\langle \Psi | H_{Int} | \Psi \rangle^{*} &=& \int\limits_{-\infty}^{\infty}\psi^{\dagger}\gamma^0
\left(-i\gamma^0\left(\frac{2M}{r}\right)^\frac{1}{2}\left(\partial_r+\frac{3}{4r}\right)\right)\psi\ d^3r\nonumber\\ 
&&+4\pi \int\limits_{0}^{\infty}\psi^{\dagger}\gamma^0 \left(-i\gamma^0\left(\partial_r \left(2Mr^3\right)^\frac{1}{2}-\left(2Mr^3\right)^\frac{1}{2}\frac{3}{2r}\right)\right)\psi\ dr \nonumber\\
&&+4\pi\ \psi^{\dagger} \left(i\left(2Mr^3\right)^\frac{1}{2}\right)\psi \mid_{r=0} \nonumber\\
&=&\langle \Psi | H_{Int} | \Psi \rangle+i\int\limits_{-\infty}^{\infty}\delta(r)\psi^{\dagger}\left(2Mr\right)^\frac{1}{2}\psi\ d^3 r.
\label{86}
\end{eqnarray}
So we have $H^\dagger \neq H$ and the energy eigenvalues need not to
be real and so we have to extend our search to complex numbers.

However if we have found the eigenstate we are mainly interested in
the real part in the context of this article. The real part of the
energy has to be smaller than the rest energy of the particle to
provide a (if we abstain from instability) stationary state. It turns out that the imaginary part is always negative.
Therefore we define

\begin{equation}
E=E_R-iE_{Im},
\label{87}
\end{equation}
with $E_{Im}>0$. 
As already mentioned in the introduction the imaginary part is connected to the instability 
of the states. This instability corresponds the analogue fact in the framework of particle mechanics that a particle which has passed the horizon cannot get back to the area outside the horizon and disappears in the singularity at the centre of the black hole. In the case of quantum mechanics this means there is an effective inward current. 
By recalling now Eqs. $(\ref{91})$ and $(\ref{92})$ and using the eigenvalue equation we obtain

\begin{equation}
| \psi (t_2) \rangle={\rm exp}\left[-iE(t_2-t_1)\right] | \psi (t_1) \rangle.
\label{93}
\end{equation}
With the help of Eq. $(\ref{87})$ we can transform Eq. $(\ref{93})$ to

\begin{equation}
| \psi (t_2) \rangle=\left\{{\rm exp}\left[-iE_R(t_2-t_1)\right] | \psi (t_1) \rangle\right\} {\rm exp}\left[-E_{Im}(t_2-t_1)\right].
\label{94}
\end{equation}
So it can be seen directly that a negative imaginary energy part results in an exponential decrease of the probability density.
This is only another way to express the fact that probability density vanishes with time. Such a phenomena can only be interpreted as the effect of a current sink at the singularity of the centre. If probability density extends to the area within the horizon it cannot escape and has to end at the singularity in the centre. Without this special structure of space time connected to black holes such a strange phenomena could not take place. This means that the decay rate of the states depends on the properties of the wave function near the origin. A small probability density near the singularity leads to a long decay time.
Thus states with a large angular momentum, i.e. a small probability density near the 
origin are very stable. This shows up explicitly in the decreasing imaginary part of the energy as a function of the 
displacement of the probability density outwards (see Figures). 
For a more extensive discussion of this decay phenomenon the reader can consult ref. \cite{Lasenby:2002mc}.

\section{Figures and Discussion}

\begin{figure}
\vspace*{0cm}
\hspace*{2cm}
\epsfig{figure=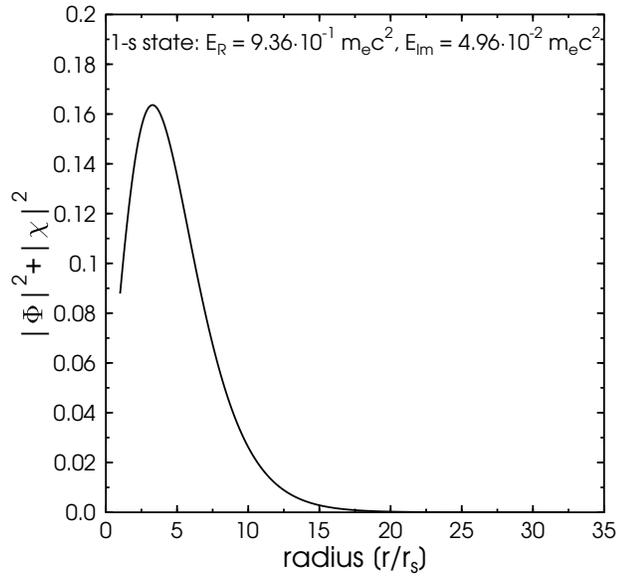,width=9.5cm,angle=0}
\vspace*{0cm} 
\hspace*{0cm}
\caption{\label{figure1}
$|\Phi|^2+|\chi|^2$ as function of $r/r_S$ for the 1s-state corresponding to an energy of $E_R=9.36 \cdot 10^{-1} {\rm m_e c^2}$, $E_{Im}=4.96 \cdot 10^{-2} {\rm m_e c^2}$.}
\end{figure}

\begin{figure}
\vspace*{0cm}
\hspace*{2cm}
\epsfig{figure=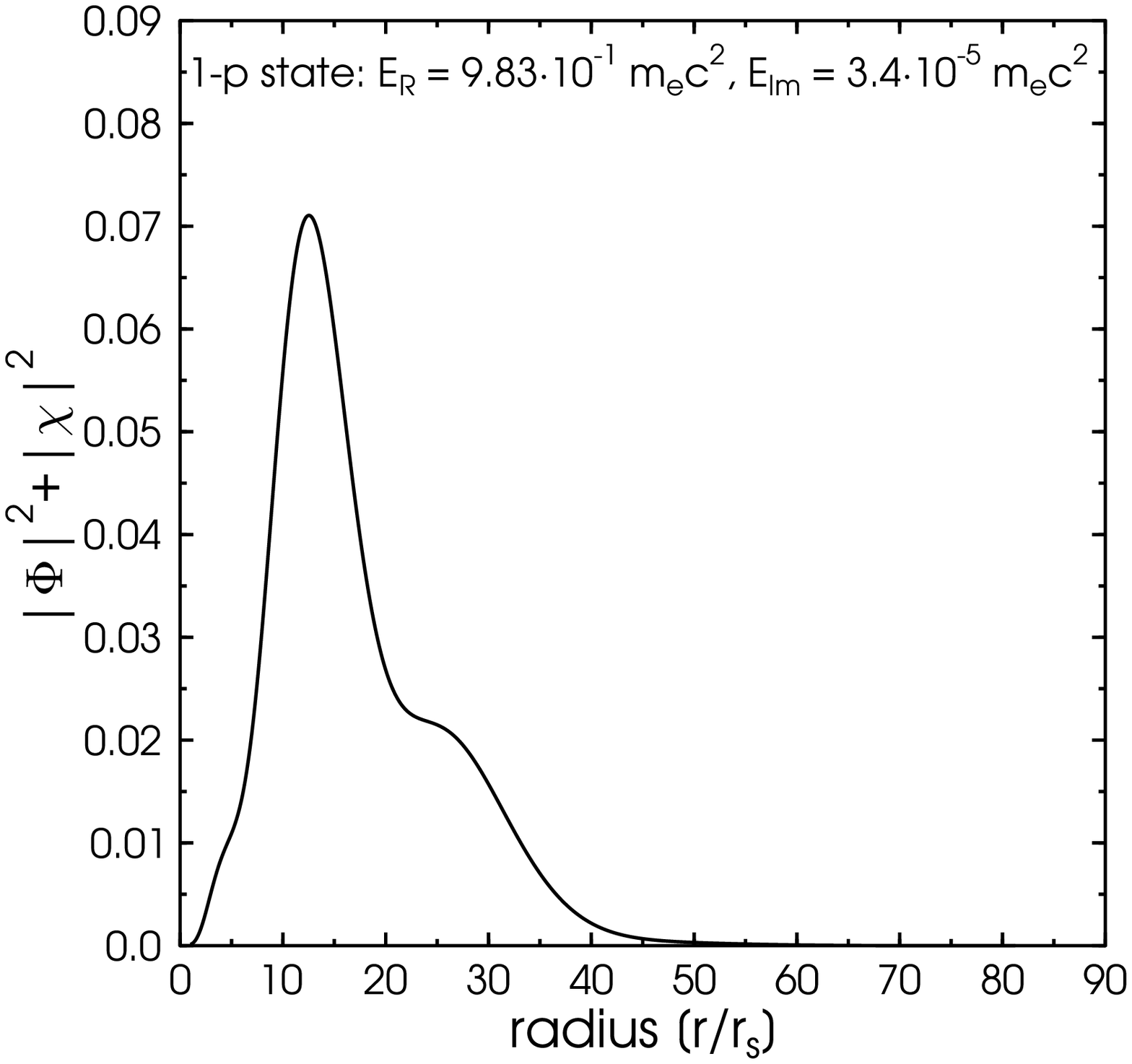,width=9.5cm,angle=0}
\vspace*{0cm} 
\hspace*{0cm}
\caption{\label{figure4}
$|\Phi|^2+|\chi|^2$ as function of $r/r_S$ for the lowest state of $J=3/2$ (1p-state) corresponding to an energy of $E_R=9.83 \cdot 10^{-1} {\rm m_e c^2}$, $E_{Im}=3.4 \cdot 10^{-5} {\rm m_e c^2}$.}
\end{figure}

\begin{figure}
\vspace*{0cm}
\hspace*{2cm}
\epsfig{figure=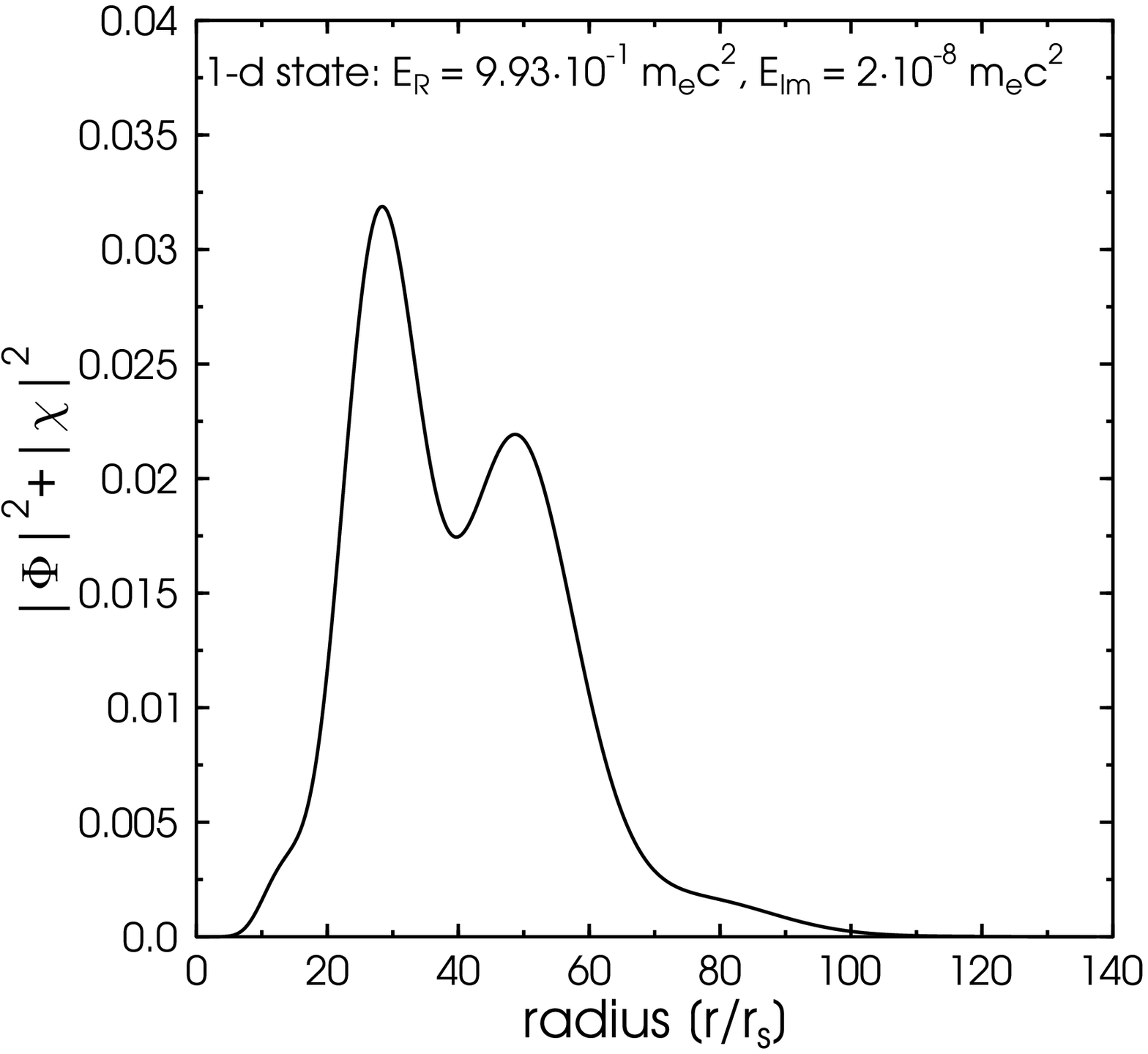,width=9.5cm,angle=0}
\vspace*{0cm} 
\hspace*{0cm}
\caption{\label{figure5}
$|\Phi|^2+|\chi|^2$ as function of $r/r_S$ for the lowest state of $J=5/2$ (1d-state) corresponding to an energy of $E_R=9.93 \cdot 10^{-1} {\rm m_e c^2}$, $E_{Im}=2 \cdot 10^{-8} {\rm m_e c^2}$.}
\end{figure}

The solutions of the Dirac equation in a Schwarzschild background differ significant from the solutions of the corresponding equation with a coupled Coulomb-potential describing a hydrogen atom. The main difference is the non-stability of the states connected to the non-Hermiticity of the Hamiltonian. In the gravitational field real bound states exist only in an approximative sense and if the angular momentum becomes large the main part of the probability density lies far away from the origin. Using the Newtonian
gravitational potential would correspond exactly to the hydrogen atom with other pre-factors. Thus the deviations in case of gravity arise from the description obtained by considering general relativity.
\newline
The other comparison concerns the relation to the orbits of classical particles. In case of classical mechanics on a Schwarzschild background there are two different kinds of states or orbits corresponding to the obtained quantum mechanical states. On the one hand there are orbits with high angular momentum preserving particles from falling into the black hole. On the other hand there are possible orbits making the particle crossing the horizon. If such a crossing of the horizon takes place the particle can never reach the area outside again and falls to the centre of the black hole. The orbits or states of the first category are stable in an absolute sense. This is different in the case of quantum mechanics where no such discrete classification can be made leading to the fact that all states have an imaginary part which never vanishes. 
\newline
Qualitatively, we might also obtain similar results with the Klein-Gordon equation since the special properties of the considered situation, e.g. the instability of the states, are independent of specific aspects of different particles. However, one does not expect to get the same energy levels because of the special aspects related to the inner degree of freedom of Fermions leading to an additional Spinor contribution to the Hamiltonian.  The additional Spinor connection term gives the reason why the energy eigenvalues for Dirac particles are different from those of Klein-Gordon particles. Besides, the description of fermions on curved space time is only possible by introducing the mathematical concept of the vierbein. In spite of the more difficult solution it is more interesting to consider the Dirac equation because most fundamental particles are fermions and thus obey the Dirac equation.  
\newline
To obtain concrete Figures and energy eigenvalues we have chosen the special case of an electron having a rest mass of
$m_e=9.109 \cdot 10^{-31} {\rm kg}$ in the gravitational field of a black hole with
mass $M=1.825 \cdot 10^{14} {\rm kg}$. To specify a particle with a certain rest mass is important since the concrete eigenvalue of the energy depends on it according to the relativistic relation between energy and momentum. 
This setting is similar to an electron in the electromagnetic field of a proton, hence a hydrogen atom. The mass of the black hole chosen here corresponds to a Schwarzschildradius of $r_s=2.7 \cdot 10^{-13} {\rm m}$. For the lowest state of $J=1/2$ (see Figure \ref{figure1}) i.e. without orbital angular momentum and the lowest states for $J=3/2$ and $J=5/2$ (see Figures \ref{figure4} and \ref{figure5}) $|\Phi|^2+|\chi|^2$ is plotted as a function of $r/r_s$. 
One can see that increasing orbital angular momentum leads to a distinct movement of the probability density outwards. In Figure \ref{figure1} the  probability density at the horizon is still more than 50 \% of the peak probability density. In the case with orbital angular momentum of 1 one observes that the contribution near the horizon nearly vanishes.

\section{Acknowledgements}

Primarily we want to thank H. St\"ocker for stimulating this very interesting project as a part of his 
lecture on Quantum Mechanics II. Without his idea to deal with such a topic in the context of a student summer project this article would not exist. Further we are grateful for inspiring and helpful discussions with A. Lasenby and S. Dolan from whose work we profited a lot.

\end{document}